# Simulation Prediction and Experiment Setup of Vacuum Laser Acceleration at Brookhaven National Lab-Accelerator Test Facility

L. Shao*, D. Cline, X. Ding, UCLA, Los Angeles, CA 90095, U.S.A

Y.K. Ho, Q. Kong, J. J. Xu, Fudan University, Shanghai, 200433 China

I. Pogorelsky, V. Yakimenko, K. Kusche, BNL-ATF, Upton, NY 11973 U.S.A.

## Abstract

This paper presents the pre-experiment plan and prediction of the first stage of Vacuum Laser Acceleration (VLA) collaborating by UCLA, Fudan University and ATF-BNL. This first stage experiment is a Proof-of-Principle to support our previously posted novel VLA theory. Simulations show that based on ATF's current experimental conditions, the electron beam with initial energy of 15MeV can get net energy gain from intense CO2 laser beam. The difference of electron beam energy spread is observable by ATF beam line diagnostics system. Further this energy spread expansion effect increases along with the laser intensity increasing. The proposal has been approved by ATF committee and experiment will be the next project.

---

*Corresponding Author. Tel.: +16313444635.
Email: leishao@ucla.edu

# Introduction

Nowadays laser technology has significantly improved in various aspects. Especially the laser energy has already reached 1MJ, with power at 100PW [1]. This soaring laser power and intensity has made Laser Driven Acceleration more and more popular in high energy area. In the recent decades a variety of laser driven acceleration topics have been developed, such as Inverse Free-Electron Laser Acceleration (IFEL) [2], Plasma Wake Field Acceleration (PWFA) [3], and Laser Wake Field Acceleration (LWFA) [4] and so on.

Vacuum Laser Acceleration (VLA) is always a controversial topic; however it has some advantages in contrast to medium-based acceleration techniques. The 1979 Lawson Woodward theorem [5] states that the free electron has no net energy exchange from infinite electric-magnetic field in vacuum. However, many assumptions of this theorem have been challenged due to the highly developed laser technology. Our group's previous work [6,7,8,9] has initially discovered that free electrons have the possibility to get net energy exchange from a tightly focused laser field and this acceleration effect increase along with increasing laser intensity. Simulation has found that with the laser intensity $a_0 \gtrsim 1$ this net energy exchange becomes obvious and observable, while the laser intensity reaches $a_0 = 3 \sim 5$ or up the acceleration effect becomes significant, which we call Capture Acceleration Scenario (CAS) [7]. When the laser intensity $a_0$ is around 100, the acceleration gradient could be very large, up to GeV/m. Theoretically, it's also proved that in a tightly focused laser beam there is a special region, where the field phase velocity is relatively low [8]. In some parts of the area the phase velocity is even lower than light speed in vacuum, $c$. This underlying physics feature makes it possible for electrons to be captured in the laser beam and accelerated violently. In this novel scheme of vacuum laser acceleration, there is no optical element required to confine interaction length.

This letter reviews the new proposed Vacuum Laser Acceleration experiment approved at BNL-ATF. And it briefs the underlying physics first, and then presents the simulation results from BNL-ATF current experimental condition; at the last describes the experiment design and plan according to simulation research with current ATF experimental condition.

## Underlying physics

We use Gaussian-Hermit (0,0) mode to describe the laser field [10]. The phase function will be:

$$\varphi(x,y,z,t) = g(x,y,z) - \omega t, \tag{1}$$

where $g(x,y,z) = kz - \varphi(z) + \frac{k(x^2+y^2)}{2\mathrm{R(z)}} - \varphi_0$ is the phase front. The phase velocity along a certain trajectory $J$ must satisfy the equation:

$$\frac{\partial\varphi}{\partial t} + \left(V_\varphi\right)_J \cdot (\nabla\varphi)_J = 0, \tag{2}$$

where $(\nabla\varphi)_J$ and $\left(V_\varphi\right)_J$ are the phase gradient and the phase velocity respectively along the electron trajectory. Therefore, we derive the phase velocity formula as:

$$V_\varphi = \omega/\nabla \mathrm{g} \cdot \boldsymbol{l}_0, \tag{3}$$

where $\boldsymbol{l}_0$ is the unit vector along the certain direction. Then we have $V_{\varphi z} = \omega/\nabla \mathrm{g} \cdot \vec{\boldsymbol{z}}$ for the phase velocity along the laser propagation direction-axis z; and $V_{\varphi m} = \omega/|\nabla \mathrm{g}|$ for the phase velocity along the gradient direction on electron trajectory during interaction. Obviously $V_{\varphi m}$ provides the lowest phase velocity. Figure 1 represent the minimum velocity $V_{\varphi m}$ distribution on $y$=0 plane. In the Figure 1 we can see that on the $z$-axis along propagation ($x$=0) direction the phase velocity is no less than $c$, especially in the central focus spot ($z$=0) the phase velocity is much higher than $c$. In this area, electrons will experience high phase slippage with the laser field and gain no energy.

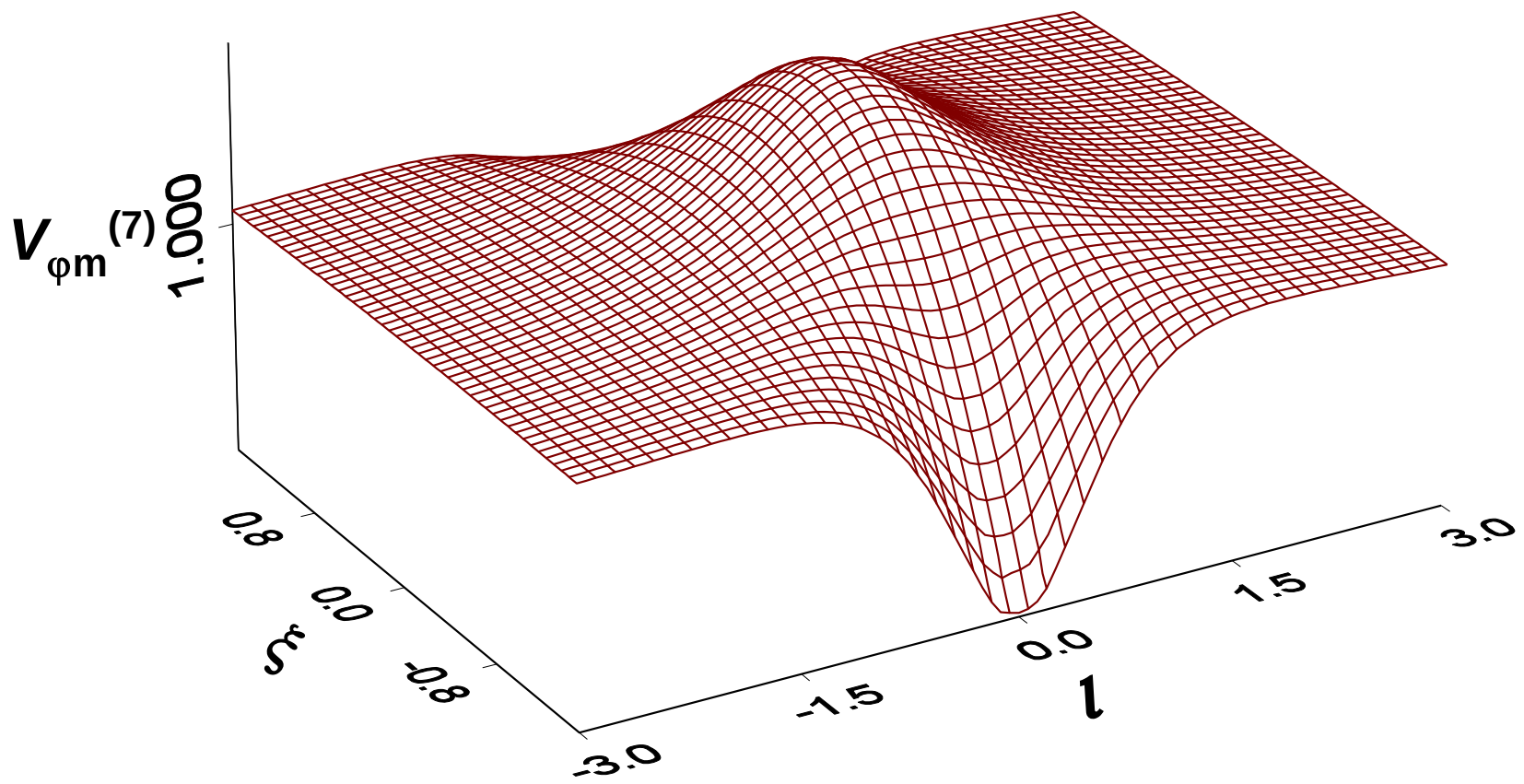

Fig. 1: The minimum phase velocity distribution on *y=0* plane, spot waist $w_0 = 30\mu m$, $\varepsilon = x/w_0$, $l = z/Z_R$, where $w_0$ is the laser waist size and $Z_R$ is the Rayleigh Length.

Also in Figure 1, however, there appear to be regions where the phase velocity is quite low, lower than *c*. These regions are located at both sides of the laser profile line *w(z)* and centered at focus spot (*x*=*z*=0) and extend a few $w_0$–length in transverse direction and a few Rayleigh Lengths in longitudinal direction. This feature exists in a focused laser beam because, unlike a plane wave, different wave fronts have different radii of curvature. In the gradient direction of phase front, the phase velocity of the laser beam reaches minimum. In equation (1), it comes from the phase factor $k(x^2 + y^2)/2R(z)$ - the diffraction effect. Therefore, if the electron is injected in a proper position, the transverse force can keep it moving along the trajectory close to the laser profile; thus the electron will experience the low phase velocity and the phase slippage would be slow enough so that electron can be captured in the acceleration phase and continue being violently accelerated.

Laser intensity is another key factor to perform this VLA mechanism. Due to the limited subluminous phase velocity distribution, high laser intensity is required to accelerate electrons in a short distance to catch up the matching phase velocity. Figure 2 presents the acceleration scale versus laser intensity with all other parameters fixed, laser spot size 30microns and initial electron energy 5MeV.

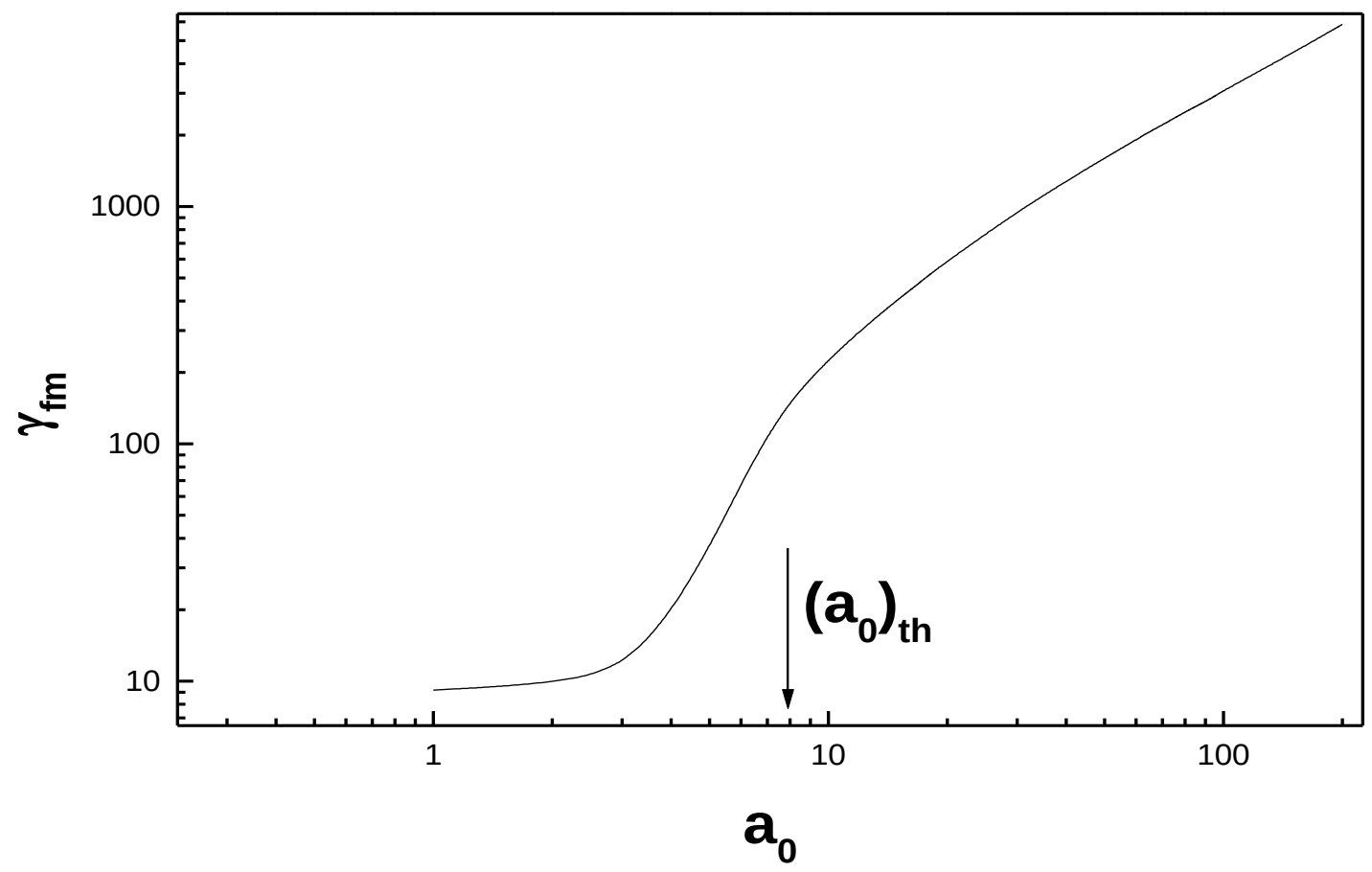

Fig. 2: Final energy vs. Laser intensity $a_0$, laser spot size $w_0 = 30\mu m$, initial electron energy 5MeV.

In Figure 2 we can see that the final energy gain of electrons increases along with increasing laser intensity. And it is found that in the range of intensity from $a_0 \approx 1$ to $a_0 \approx 5$ the net energy exchange scale is nonlinear to the laser intensity. The net energy exchange significantly increases when laser intensity reaches around $a_0 \approx 5$. Electrons can be captured in the acceleration phase and keep continuously being accelerated. We named this phenomenon as Capture and Acceleration Scenario (CAS) – an extreme case of VLA. In our previous work we pointed out that $a_0 \approx 5$ is the threshold of laser intensity to perform CAS. Beyond $a_0 \approx 5$, in CAS section, the energy exchange is linear to the laser intensity [7]. This promising project has high practical application value for new generation of accelerator.

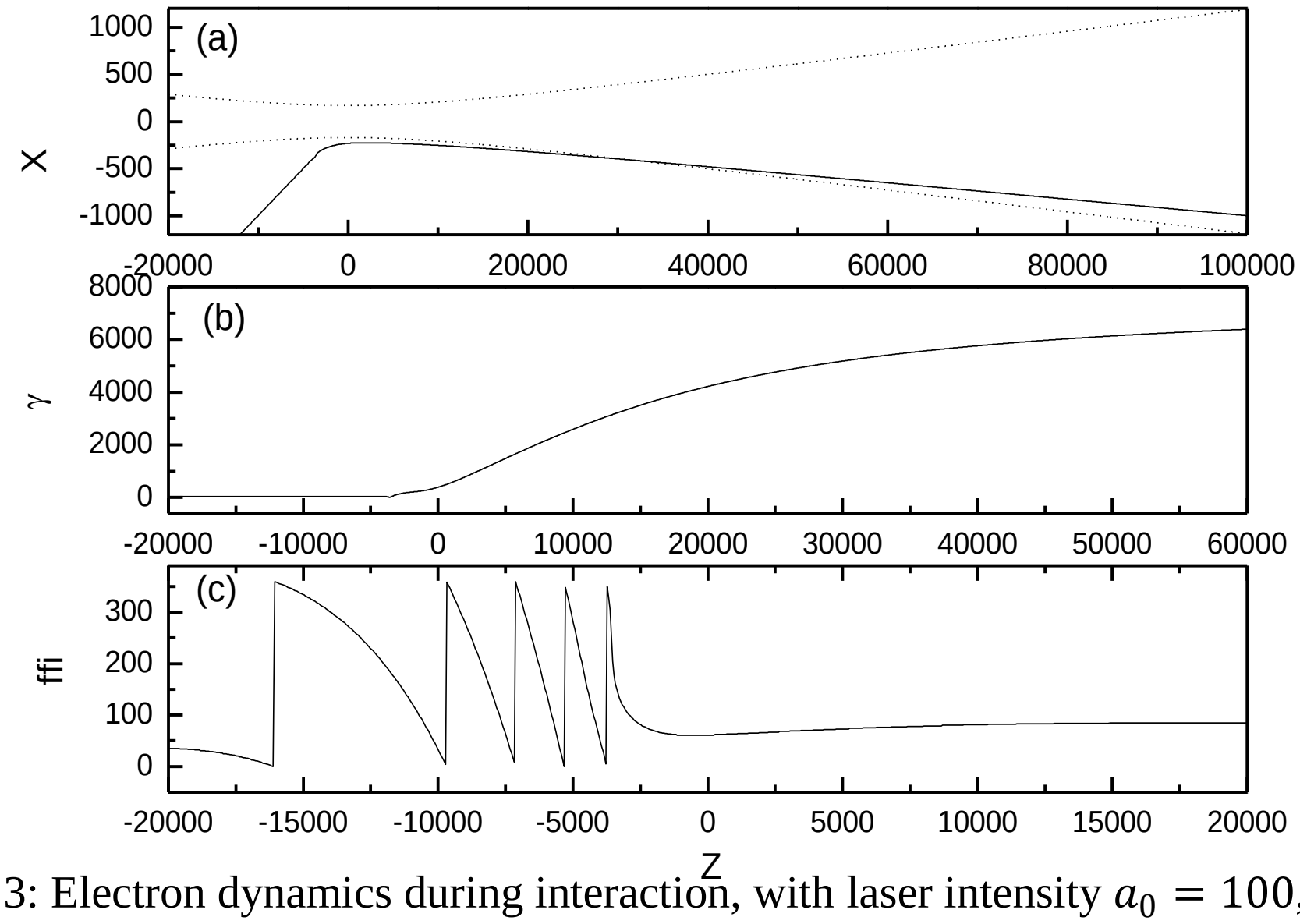

Fig. 3: Electron dynamics during interaction, with laser intensity $a_0 = 100$, other parameters are the same as Fig. 2. (a) is the single electron trajectory in *x*-*z* plane; (b) is the energy change; (c) is the phase slippage that electron experiences during interaction.

Figure 3 is the detailed dynamics of one typical case of single electron from Figure 2 at $a_0 = 100$. Figure 3(a) shows the electron trajectory on *x*-*z* plane. The electron is not scattered off the laser field immediately, instead, the electron is captured in the laser profile and moves along with the laser. This is new and different from any of Elastic-Scattering or Inelastic-Scattering. Figure 3(b) represents the electron energy change versus time. This is consistent with Figure 3(a). After the electron moves through the focus spot position on *z*-axis, the electron is captured in the acceleration phase and continues to be accelerated during most of the entire interaction. The final energy gain is up to GeV with such intense laser. Recalling the subluminous phase velocity feature we discussed in the above, Figure 3(c) gives the phase that the electron experiences along its trajectory. Starting from the same position, the laser focus spot, the phase slippage slows down considerably, and electron almost stays in the same phase while propagating with the laser beam.

Thus if the electron catches the acceleration phase when it enters the laser beam and gets violent acceleration at the moment, it could stay in the acceleration phase and continue being accelerated.

# VLA Prediction with ATF Current Experimental Conditions

The current ATF's CO2 laser system delivers peak energy at 5J; pulse length 5ps, 1ps, 500fs; focus spot around 50μm to beam lines in experimental hall. Therefore the laser intensity could be $a_0 \approx 0.9 \sim 2.2$. Such laser intensity level is not sufficient for CAS; however it could perform the obvious net exchange. This is our first stage of VLA experiment – proof of principle.

ATF's electron beam system provides a high quality electron beam with low emittance and it routinely operates at energy above 40 MeV to avoid strong space-charge effects. ATF has two SLAC-type S-band linac sections. Basically there are two solutions to obtain a lower-energy beam: one is to adjust the first linac section phasing in acceleration with a larger accelerating gradient but phasing the second linac section in deceleration to obtain a lower energy; and the other is to adjust both linac sections phasing in acceleration but with a lower accelerating gradient. PARMELA code was used first to simulate beam energy spectrums for both solutions at 15MeV. The second solution can offer a very small energy spread, 0.1%, while the first one offers a large energy spread. And it was successfully tuned to the end of the beam line at 15MeV beam energy. The achieved beam test results at 200 pC are summarized in the following table 1:

| Laser: | | | e-beam: | |
|---|---|---|---|---|
| Spot Size $w_0$ | $30\mu m$ | | Initial Energy $E_0$ | 15MeV |
| Wave Length | $10.6\mu m$ | | Initial Emittance | ~1.0 $mm\cdot mrad$ |
| Intensity $a_0$ | 0.9, 1.5, 2.2 | | $\Delta E_i/E_i$ | $\sim\times 10^{-3}$ |
| | | | beam size (at focus spot) | $168\mu m$, $200\ \mu m$ |

Table 1: Practical parameters at ATF, used for simulation.

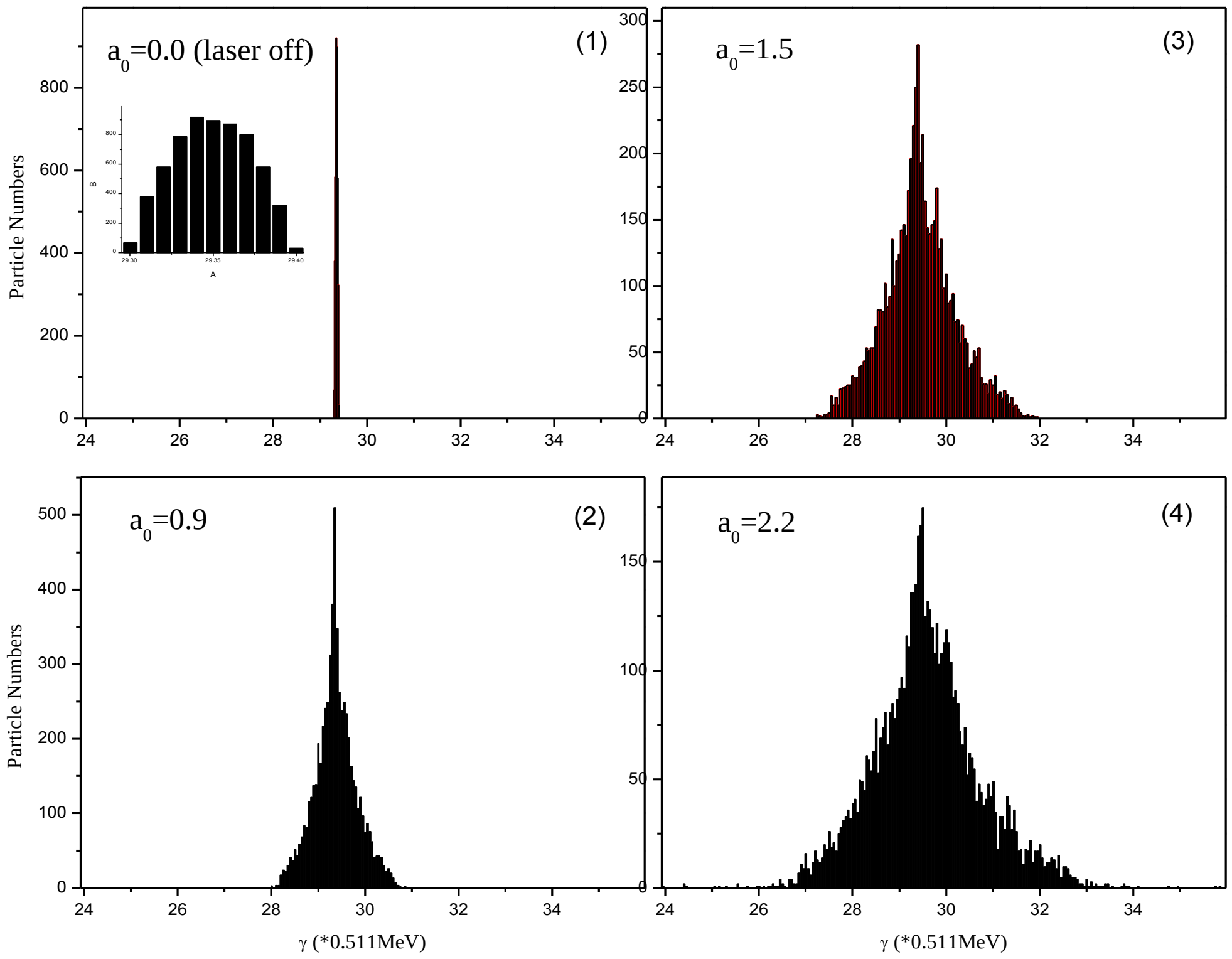

Fig. 4: e-beam energy spread distribution before and after interacting with laser beam of different intensities; other parameters are $w_0 = 30\mu m$; initial e-beam energy 15MeV, emittance 1.2mm·mrad.

In simulation, the electron beam enters along the same direction as the laser beam on $z$-axis and encounters laser beam at the laser focus spot. The simulation results are shown in Figure 4. We only show the simulation of e-beam 168microns. Figure 4 represents the final electron beam energy spread after interacting with different laser intensity. Figure 4(1) is the case of laser off; the electron beam doesn't experience any field. Figure 4(2), 4(3) and 4(4) represent the results of interacting with laser intensities of $a_0 = 0.9$, $a_0 = 1.5$ and $a_0 = 2.2$ respectively.

We use Gaussian regression to analyze the energy distribution. By comparing Figure 4(1) and Figure 4(2) we can observe that the electron beam energy spread, $\sigma$ of Gaussian distribution, expands to $10^{-2}$ from the initial $10^{-3}$. ATF's current spectrometer and diagnostic system can distinguish 0.1% accuracy and tell 0.05% accuracy. The changes of energy spread between "laser on" and "laser off" can be measured by the spectrometer on beam line. The original e-beam is at 15MeV and ~0.1 % energy spread, which is about ~±15keV. And the final e-beam energy spread is centered around 15MeV as well; however the energy spread increases to around ~3.7%, which gives about ~±550keV. This is ten times bigger than the energy spread with "laser off" case. The maximum energy loss and gain could be ~±1.5MeV. In Figure 4(3) and 4(4) the energy exchange effect is more significant while the laser intensity increases. This is consistent with our theory and proves the signal of net energy exchange coming from laser acceleration mechanism. In Figure 5 it can be seen the trend of the acceleration effect versus laser intensity.

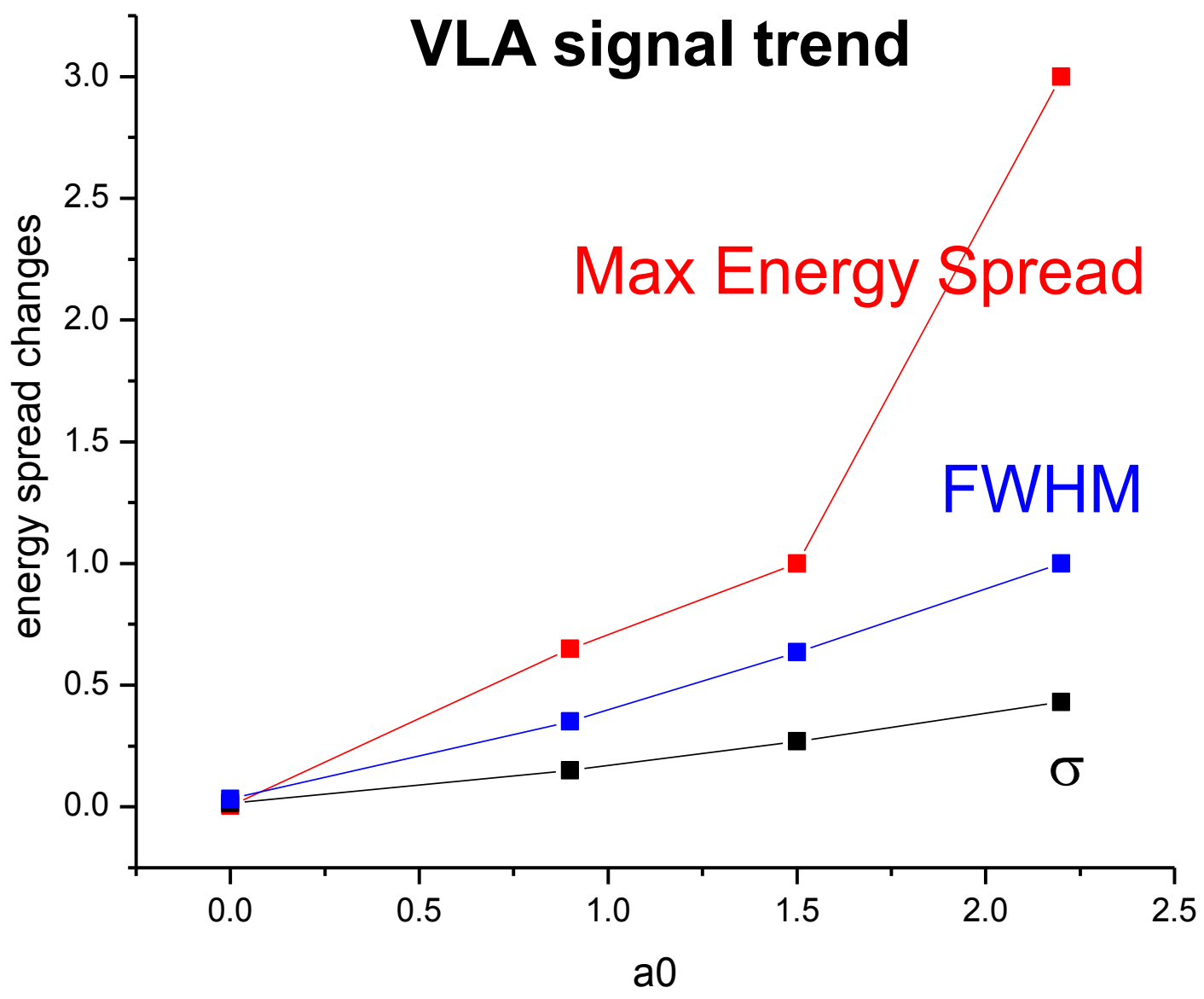

Fig. 5: The trend of energy spread changing versus laser intensity $a_0$, the same result from simulation in Figure 4.

From Figure 4 and Figure 5 we can see with laser intensity $a_0 = 2.2$ the maximum energy exchange can be up to 3.0MeV with initial 15MeV.

This proof-of-principle of VLA is promising and important. With this novel VLA mechanism, electrons can be accelerated in very short distance with high acceleration gradient. It could be a new generation of desktop acceleration. In the following we describe the experiment design and setup at BNL-ATF.

## Experiment Design, Setup and Diagnostic System

This proof-of-principle experiment is the first stage of project. And it is assigned on beam line #1 at BNL-ATF. We expect to see the signal of the energy spread expansion shown by simulation in the above. Figure 6 is the layout of the experiment design. Target parts (3) and (4) in Figure 6 are the pinhole and Germanium plate respectively. The pinhole is used for electron-beam and laser beam alignment. On beam line #1 at ATF, one He-Ne laser is used to simulate electron beam transporting and another He-Ne laser is used simulate the CO2. With the iris at the downstream position, we use pinhole and these two local He-Ne lasers to achieve alignment. The Germanium plate is used for synchronization. An electron-beam going through Germanium plate will generate plasma, which blocks CO2 laser beam. By measuring the signal on the detector (7) we can achieve synchronization to ps. After alignment and synchronization are achieved the optics setup outside vacuum chamber will be disabled for interaction measurement. The yellow parallel lines denote the initial incoming CO2 laser beam. The CO2 laser enters from the downstream window and then goes backward along the beam line by being reflected by the downstream flat mirror (1) in the chamber. And parabolic mirror (2) will focus the incoming CO2 laser at the interaction spot and reflect the CO2 laser again to propagate forward the same direction as electron beam. After interacting, the e-beam is directed out of the interaction vacuum chamber downstream on beam line #1. There is a dipole positioned right after the chamber downstream can bend the e-beam by 90° degrees to guide the output electron-beam to the spectrometer which then can measure the energy spread of the output e-beam. The major interaction area is only about several centimeters (a few Rayleigh lengths).

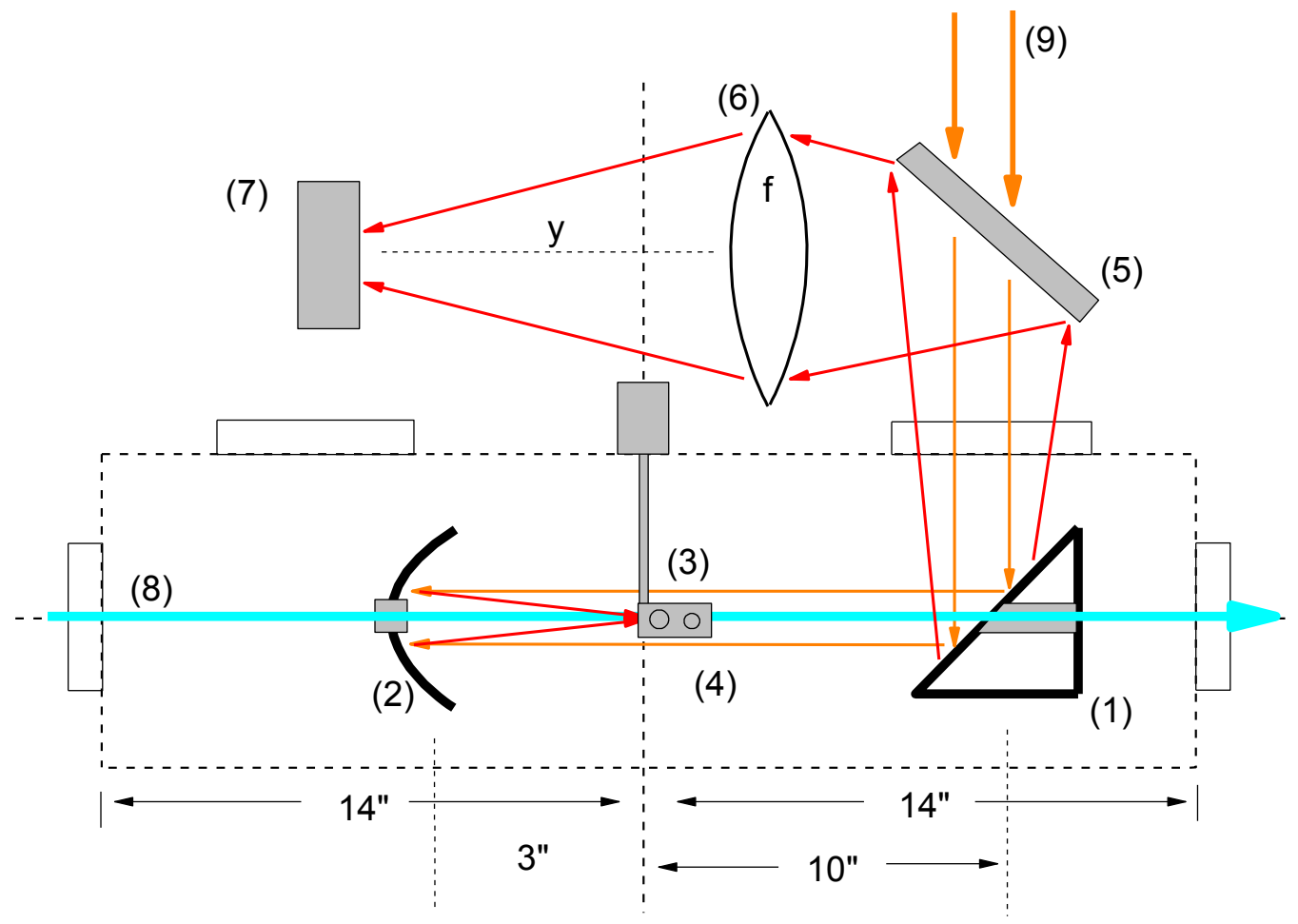

Fig. 6: Experiment design and diagnostic system. It will be on Beamline #1, (1) Flat mirror #1 with small hole, reflecting CO2 laser; (2) Focus parabolic mirror #2 with small hole, short focus length f=3”; (3) and (4) The pinhole and Germanium plate for alignment and synchronization; (5) 45 °Beam splitter for alignment and diagnostic; (6) and (7) outside vacuum chamber detector setup for beam synchronization; (8) electron beam; (9) CO2 beam.

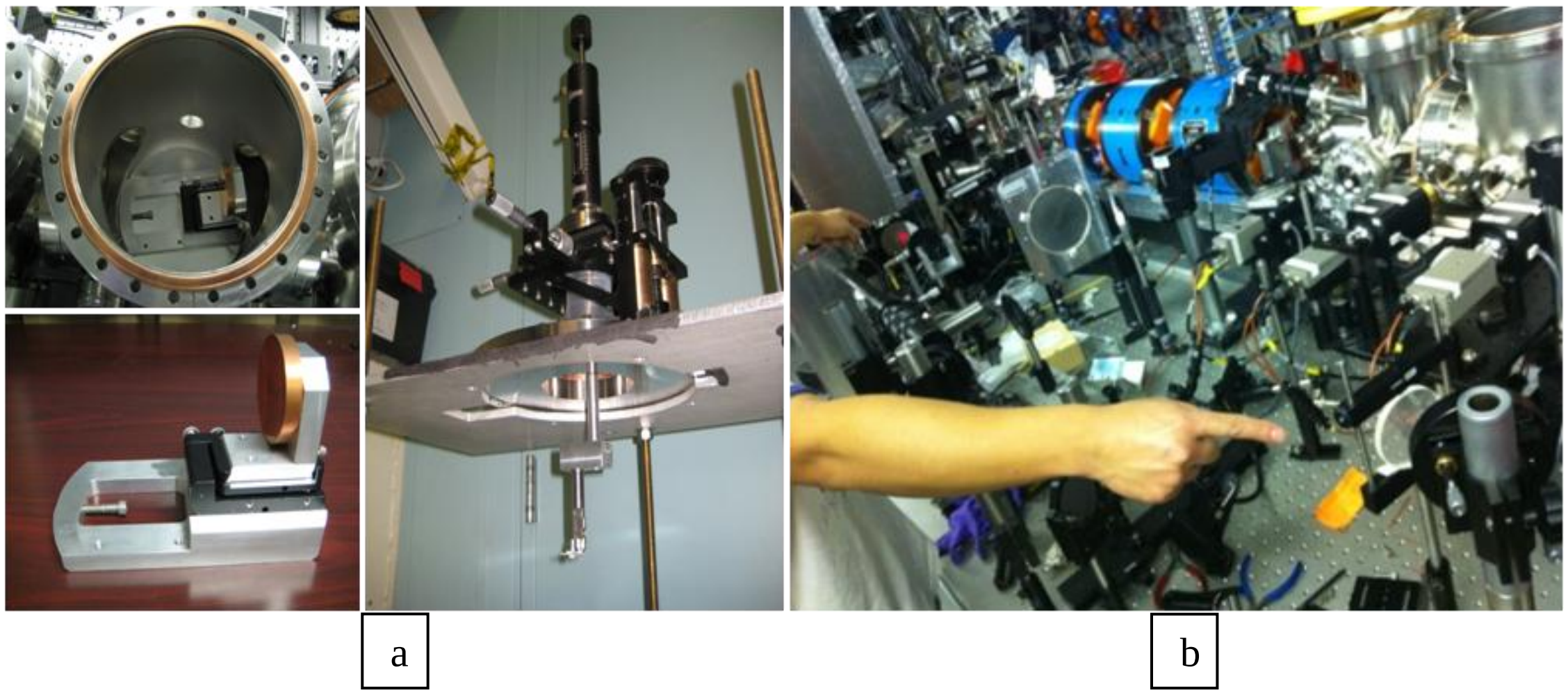

Fig. 7: Experiment devices and setups. Upper-left in (a) is the vacuum chamber for interaction; the lower-left in (a) is the parabolic mirror with 5mm hole for electron-beam transporting, with focus length 3”, sitting in the chamber (upper-

left in (a)); the right side in (a) is the target of pin-hole and Germanium plate mounted on the flange, which will be installed in the vacuum chamber; (b) is the outlook of beam line #1 and optics setup table.

Figure 7 shows the experiment devices, set up on beam line #1. The upper-left in Figure 7-(a) is the vacuum chamber for interaction. The lower-left in (a) is the parabolic mirror with a 2mm hole, used for reflecting and focusing laser. The parabolic mirror sits in the vacuum chamber as we can see in the upper-left in Figure 7-(a). The right side of Figure 7-(a) is the target of pin-hole and Germanium plate mounted on the flange which will be installed in vacuum chamber in the upper-left in Figure 7-(a). And the flange can adjust the target (the pin-hole and Germanium) in three directions. Pin-hole will be used for alignment and Germanium will be used for synchronization. The 3-D adjustment of the flange will be used to control the shifting. And the whole target will be completely retracted during experiment, after alignment and synchronization. The parabolic mirror is mounted on an adjustable stage, facing the direction of downstream. As we described in Figure 6, the incoming CO2 laser will be opposite of beam line transporting direction. Parabolic mirror will reflect CO2 back to guide it to the same direction of beam line transporting and focus CO2 in 3”, the position of the target. The 2mm hole in the parabolic mirror will be set to the same height as e-beam in order to let e-beam go through the mirror. Figure 7-(b) shows the outlook of beam line #1 setup and optics setup for guiding CO2 laser outside the chamber.

# CONCLUSION

Based on the theoretic research and computer simulations, vacuum laser acceleration can be achieved under current BNL-ATF experimental conditions. The signal is not significant; however it is obvious enough to show a proof-of-principle. And it will be the first time to prove and implement real Vacuum Laser Acceleration. This is our first stage of Vacuum Laser Acceleration.

Based on this principle-of-proof experiment result at BNL-ATF, the stage-2 of this novel VLA mechanism experiment is planning to collaborate with Lawrence Livermore National Lab and utilize their extreme powerful laser platform to prove CAS-the continuous violent acceleration to GeV.

This new concept of Vacuum Laser Acceleration will be a remarkable discover in high energy physics and will result in tremendous applications in various fields.

## Acknowledgements

This work was supported by the U.S. Department of Energy under award numbers DE-FG02-92ER40695 (UCLA). And thanks to the collaboration and support from Fudan University, Shanghai China, and Brookhaven National Lab-Accelerator Test Facility, Upton USA.

## References

[1] A. Herller, ***Science & Technology Reiview***, 25, (2000)

[2] R.B. Palmer, ***J. Appl. Phys. 43***,3041(1972)

[3] Schoessow，1994，***Advanced Accelerator Concepts, AIP Conf***. Proc. No 335, AIP

[4] F. Amiranoff, S. Baton, D. Bernard, *et al.*, ***Phys. Rev. Lett.,*** 81, 995-998 (1998)

[5] K. T. McDonald, ***Phys, Rev. Lett***. 80, (1998) 1350

[6] J. X. Wang, Y. K. Ho, *et al.*, ***Phys. Rev. E***, 60, 7473, (1999)

[7] J. Pang, Y. K. Ho, N. Cao, L. Shao, *et al.*, ***Appl. Phys. B***, 76, 617-620, (2003)

[8] J. Pang, Y. K. Ho, X. Q. Yuan, *et al.*, ***Phys. Rev. E***, 66, 066501, (2002)

[9] L. Shao, D. Cline, *et al.*, ***2005 Particle Accelerator Conference, AIP Conf.***, 2959 (2005)

[10] O. Svelto, D. C. Hanna, ***Principles of Lasers***, 3rd ed. (Plenum, New York, 1985)